\documentclass[bibyear]{aa}

\usepackage{graphicx}
\usepackage{amsmath}
\usepackage{epsfig}
\usepackage{txfonts}

\begin{document}

\def\teff{$T_{\rm eff}$}
\def\logg{$\log g$}
\def\micro{$\xi$ }
\def\kms{km\,s$^{-1}$}
\def\p{$\pm$}
\def\vsini{$v\sin i$}

\newcommand\he[2]{He\,\textsc{#1}\,$\lambda$\,{#2}}
\newcommand\n[2]{N\,\textsc{#1}\,$\lambda$\,{#2}}
\newcommand\car[2]{C\,\textsc{#1}\,$\lambda$\,{#2}}

   \title{WR\,138: new results from X-ray and optical spectroscopy\thanks{Based on observations collected at the Observatoire de Haute Provence (France), the San Pedro M\'artir observatory (Mexico), and with {\it XMM-Newton}, an ESA science mission with instruments and contributions directly funded by ESA member states and the USA (NASA).}}

   \author{M. Palate\inst{1}
          \and
          G. Rauw\inst{1}
          \and
          M. De Becker\inst{1}
          \and
          Y. Naz\'e\inst{1}\thanks{Research Associate FRS-FNRS.}
          \and
          P. Eenens\inst{2}
          }

   \institute{Institut d'Astrophysique et de G\'eophysique, 
   Universit\'e de Li\`ege, B\^at. B5c, All\'ee du 6 Ao\^ut 17, 4000 Li\`ege, Belgium\\
   \and
   					Departamento de Astronomia, Universidad de Guanajuato, Apartado 144, 36000 Guanajuato, GTO, Mexico
             }
  \offprints{M.\ Palate}
	\mail{palate@astro.ulg.ac.be}

   \date{Received 28 June 2013 / Accepted 8 October 2013}

	\abstract
   {Massive-binary evolution models predict that some systems will go through an evolutionary phase where the original primary has become a supernova and left a compact object behind that then orbits a Wolf-Rayet (hereafter, WR) star. WR\,138 is an X-ray bright WR star that has been described as a triple system, including a compact companion in a short-period orbit.}
   {Our goal is to search for spectroscopic evidence of a compact companion around WR\,138.}
   {We used optical and X-ray spectra to search for signatures of a compact companion, which can be revealed by systematic variations in WR optical spectral lines induced by orbital motion of the compact companion or by hard, luminous X-rays from accretion onto this companion.}
   {The optical spectra display emission-line profile variations that are most probably caused by clumps inside the stellar winds. The radial velocities do not vary on a short time-scale compatible with the suggested orbital period of a putative compact companion. The X-ray spectra are found to be normal for a WN5-6+OB system with no indication of accretion by a compact companion.}
   {There is no evidence for the presence of a compact companion, and we therefore conclude that WR\,138 is a normal long-period (P$\sim$1521 d) eccentric WR+OB system.}

	\authorrunning{Palate et al.}
	\titlerunning{Revisiting WR\,138}

   \maketitle

\section{Introduction}
Whilst absorption lines in the spectrum of WR\,138 were recognized several decades ago, the multiplicity of this star has been an open question for a long time. Massey (\cite{Massey}) found no evidence for significant ($K \geq 30$\,km\,s$^{-1}$) radial velocity variations that would have been attributable to an orbital motion of the emission lines and suggested that the broad ($v\,\sin{i} \simeq 500$\,km\,s$^{-1}$) absorption lines were intrinsic to the WN star. 
From a series of photographic spectra, Lamontagne et al.\, (\cite{Lamontagne}) noted low-amplitude radial velocity variations of the \n{iv}{4058} line on a period of 2.3238\,d, although other periods, such as 0.39\,d, could not be fully excluded. These variations were not seen in the \he{ii}{4686} line. The absorption lines did not display variations on this period either. In addition, Lamontagne et al.\ (\cite{Lamontagne}) noted that the radial velocities (RVs) of both the emission and absorption lines vary on a long period of about 1763 or 1533\,d. As a result, they classified WR\,138 as a triple system consisting of a WN6 star orbited by an unseen low-mass companion, probably a neutron star, in a 2.3238\,d orbit and an O-type star in a 1763\,d orbit. 
The interpretation of the short-period variations as the orbital period of a compact companion was questioned by Vreux (\cite{Vreux}), who drew attention to the ambiguity of the period determination due to the aliasing problem. He suggested that alternative mechanisms, such as non-radial pulsations, could produce modulations of the spectra of WR stars on periods that could be aliases of those reported in the literature. 
Annuk (\cite{Annuk}) analysed a series of medium-dispersion spectra of WR\,138 and reported RV variations with a period of $1538 \pm 14$\,d (although a period of $1420$\,d could not be fully excluded), an amplitude of about 60\,km\,s$^{-1}$, and a moderate eccentricity of $0.3$. He was unable, however, to confirm the short-period variations of the RVs reported by Lamontagne et al.\ (\cite{Lamontagne}). 
X-ray observations are expected to provide an independent indication about the presence or absence of a compact companion in WR\,138. Indeed, the presence of a neutron star in a close orbit around a WN star should lead to substantial accretion by the former and would thus produce a strong and hard X-ray emission, as seen in high-mass X-ray binaries (see e.g. Negueruela \cite{Negueruela}). The X-ray emission from WR\,138 was detected with the {\it EINSTEIN} observatory (Pollock \cite{Pollock}) and within the All Sky Survey of {\it ROSAT} (Pollock et al.\ \cite{RASS}). However, these observations did not provide the  level of detail required to clarify the nature of the X-ray emission. In this paper, we present new X-ray observations of WR\,138 augmented by several new series of optical spectra to re-investigate whether there is a compact companion in a short-period orbit around the WN star.

\section{Observations}\label{sec:2}

	\subsection{X-ray observations} \label{sect2.1}
WR\,138 was observed serendipitously with {\it XMM-Newton} (Jansen et al.\ \cite{Jansen}) in May 2011, during three exposures of 20\,ks each centred on the massive binary HDE\,228766 (Rauw et al.\ in preparation; ID 067048). EPIC instruments (Turner et al.\ \cite{MOS}, Str\"uder et al.\ \cite{pn}) were operated in full-frame mode and the medium filter was used to reject optical and UV photons. 

The raw data were processed with SAS software version 12.0. The first two observations were affected by background flares, and we therefore rejected these high-background episodes from further processing. Because WR\,138 was located 9' off-axis in these pointings, the source sometimes fell on a gap between the detector chips or on the dead CCD of the MOS\,1 detector. As a consequence, not all EPIC instruments provided useful data for WR\,138 for each exposure. In particular, we decided to restrict our analysis to MOS\,2 and pn data. The extraction zones consist of a circle of 40'' centred on the source and a nearby circular zone of 35'' radius in a source-free region for the background. The spectra were binned with a minimum of 25 counts per bin of energy channel and a maximum oversampling of the resolution element by a factor 5.

\begin{table}
	\caption{Journal of the {\it XMM-Newton} observations}
	\begin{center}
		\begin{tabular}{c c c c c}
		\hline
		Date & Date & \multicolumn{2}{c}{Useful exposure time} \\
					&HJD$-$2450000 & MOS\,2 & pn \\
     (dd-mm-yyyy) & (ks) & (ks)  \\
		\hline
		05-05-2011& 5686.713 & 21.6 & 14.1 \\
		13-05-2011& 5694.768 & 19.9 & 13.0 \\
		18-05-2011& 5700.152 & 23.8 & 18.0 (gap)  \\
		\hline
	\end{tabular}
	\tablefoot{The date of the observation is given at mid-exposure.}
	\end{center}
\end{table}

Several other X-ray observations were also available in the HEASARC archives. WR\,138 was observed with {\it ROSAT} in October 1993, in June 1994 with the PSPC instrument, and in November 1994 with the HRI instruments (SEQ ID RP500248N00, RP500248A01, and RH500341N00). The exposure durations were 3.6, 4.2, and 42.8\,ks, respectively. WR\,138 was also observed with {\it SWIFT} in March 2011 (OBS ID 00040153001) with an exposure time of 1\,ks. Finally, a {\it CHANDRA} observation of 69.3\,ks made in January 2010 (OBS ID 11092) was also available. In all the {\it ROSAT} and {\it CHANDRA} observations, WR\,138 is located far off-axis, leading to a severe degradation of the PSF.
We downloaded the processed data and extracted the spectra of WR\,138 for these five archival observations. The extraction zone for the {\it SWIFT} data consists of a circular zone with a radius of 50'' around the star and the surrounding annulus with an outer radius of 213'' for the background. The extraction zone for the {\it CHANDRA} data is an ellipse of 33''$\times$47'' semi-axes for the source and a nearby circle of 43'' radius for the background. The {\it ROSAT} HRI and PSPC extraction zones consist of circles of 1' and 3.5', respectively, centred on the source and surrounding annuli with outer radii of 2' and 5' for the background. The minimum numbers of counts per bin for the {\it CHANDRA} and {\it ROSAT} mission are 10 cts/bin and 15 cts/bin.

	\subsection{Optical spectroscopy}
WR\,138 was monitored with the Aur\'elie spectrograph (Gillet et al.\ \cite{Gillet}) at the 1.52\,m telescope of the Observatoire de Haute Provence (OHP) during two observing campaigns in September 2011 and June 2012. The 2011 data were taken with a 1200 lines\,mm$^{-1}$ grating blazed at 5000\,\AA\ and cover a wavelength domain from 4453 to 4675\,\AA\ with a resolving power of $\simeq 20000$. The 2012 spectra were obtained with a 600 lines\,mm$^{-1}$ grating also blazed at 5000\,\AA. The wavelength domain covered by these data extends from 4448 to 4886\,\AA\ with a resolving power of $\simeq 10000$. The detector was an EEV 42-20 CCD with $2048 \times 1024$ pixels. The data were reduced in the standard way (see Rauw et al.\ \cite{bd60}) using the MIDAS software provided by ESO. The optical spectrum of WR\,138 consists of many strong and broad emission lines with only very few line-free regions. We normalized the spectra of each campaign using a number of (pseudo) continuum windows. Given the different spectral ranges of the two datasets and especially in view of the narrow spectral range of the 2011 data, it was not possible to normalize both datasets consistently. Nevertheless, within a given dataset, the normalizations are consistent, allowing us to search for line-profile variability. 
In the time of August to September 2012, we furthermore collected a series of \'echelle spectra of WR\,138 with the ESPRESSO spectrograph mounted on the 2.12\,m telescope at the Observatorio Astron\'omico of San Pedro M\'artir (SPM) in Mexico. These data cover the spectral region from 3900 to 7200\,\AA\ with a resolving power of $\simeq 18000$. The integration times of individual spectra was 15\,min. The data were reduced using the IRAF \'echelle packages. Due to the strongly peaked blaze of the ESPRESSO spectrograph and the broad spectral lines of the Wolf-Rayet star, the normalization of the spectra was very difficult. Therefore, we used the SPM data mainly for measuring RVs and for classification purposes. 

\section{Optical spectra}

\begin{figure}[ht!]
				\resizebox{\hsize}{!}{\includegraphics{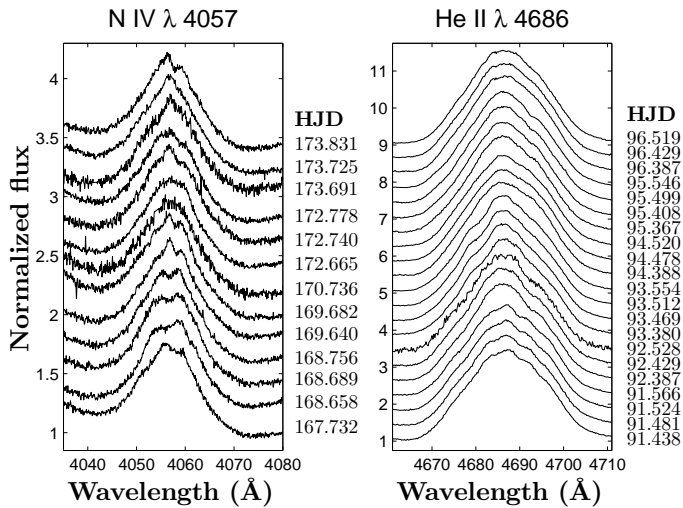}}
			\caption{Line-profile variations for the SPM (\textit{left}) and OHP\,2012 (\textit{right}) observations. The observations are vertically shifted by 0.2 (\textit{left}) and 0.4 (\textit{right}) continuum units for clarity (time is given as HJD - 2456000).}
			\label{fig:linesvar}
		\end{figure}

The OHP spectra display several emission lines: N\,\textsc{iii}\,$\lambda\lambda$ 4510-4534 and 4634-4640; N\,\textsc{v}\,$\lambda\lambda$ 4604 and 4620;  He\,\textsc{ii}\,$\lambda\lambda$ 4542, 4686, 4859, and H\,$\beta$. They also display one absorption line: \he{i}{4471}. We first performed a temporal variation spectrum (TVS, see Fullerton et al.\ \cite{Fullerton}) analysis on these spectra and found that three regions were variable: the N\,\textsc{iii}\,$\lambda\lambda$\,4510-4534 complex, the \he{ii}{4686} line, and the \he{ii}{4859} + H\,$\beta$ blend. The \he{ii}{4686} line presents the strongest variations, especially in the core of the line. These variations are clearly visible and take the form of changing subpeaks in the line core with a timescale of a few hours to days, as can be seen in Fig.\ref{fig:linesvar}. The H\,$\beta$ region displays the weakest variation, which is at the limit of the significance level (at 1\%). The variations of these emission lines are reminiscent of those of other Wolf-Rayet stars. They are commonly attributed to wind inhomogeneities (clumps) and more specifically to the statistical fluctuation of the number of clumps emitting at a specific line-of-sight velocity (see e.g. L\'epine et al. \cite{Lepine, Lepine99}).

The SPM spectra were mainly used for RV measurements and spectral-type determination. We studied the same emission lines as in the OHP data plus the \n{iv}{4058} and He\,\textsc{ii}\,$\lambda\lambda$\,4100 and 4200 lines, as well as two more absorption lines (He\,\textsc{i}\,$\lambda\lambda$\,4143 and 4921). The lines in the SPM spectra also present line profile variations that are clearly visible in Fig\,\ref{fig:linesvar}. We determined the spectral type of WR\,138 using the criteria of Smith et al. (\cite{Smith}). We found that the peak-over-continuum ratios of \n{v}{4604} over \n{iii}{4640}, \he{ii}{5411} over \he{i}{5875}, \car{iv}{5808} over \he{ii}{5411}, and \car{iv}{5808} over \he{i}{5875} are equal to $\sim1.05$, $\sim1.57$, $\sim0.80$, and $\sim1.25$, respectively, which corresponds to a spectral type between WN5 and WN6. Annuk (\cite{Annuk}) and Lamontagne et al. (\cite{Lamontagne}) found a WN6 spectral type using the old criterion of Smith (\cite{Smith68}). Adopting this old criterion, we also derive a WN6-type. We therefore classify WR\,138 as a WN5-6 star. Annuk (\cite{Annuk}) suggested an O9-9.5\,I-II type for the companion, with a high rotational velocity \vsini $\sim 500$ \kms. This high rotational velocity would be unusual for a bright giant or supergiant in a well-detached binary system. Unfortunately, we cannot refine this classification because our data do not allow us to clearly identify a sufficient number of lines belonging to the OB companion.

\subsection{Radial velocities}

As mentioned above, several scenarios have been proposed concerning the multiplicity of WR\,138. Massey (\cite{Massey}) favoured a single-star scenario, whilst Lamontagne et al. (\cite{Lamontagne}) suggested that it might be a triple system with a WR + O system in a wide orbit and a compact object closely orbiting the WR star. Finally, Annuk (\cite{Annuk}) concluded that the system consists of only of a WR + O pair in a wide orbit. To determine whether a compact companion is present close to the WR star, we performed an RV analysis on the basis of several lines. We used the above-mentioned spectral lines except for the N\,\textsc{iii} lines and \n{v}{4620}, which were too blended to provide accurate measurements. We stress that the lines are very broad and shallow and therefore lead to quite large uncertainties on the RVs. Moreover, as pointed out above, several lines present profile variations that make the RV measurement more difficult. Using the MIDAS software, we fitted Gaussians to the lines, starting from various heights above the continuum, to determine their centroid, and for this we computed the corresponding RVs. The RVs quoted in the present paper are expressed in the heliocentric standard of rest (by applying appropriate RV corrections). The RVs reported in Tables \ref{tab:vradohp11}, \ref{tab:vradohp12}, \ref{tab:vradspm12e}, and \ref{tab:vradspm12a} correspond to the mean value obtained for each series of measurements performed for each spectral line\footnote{The bluest and reddest measurements yielded the error interval quoted in Tables \ref{tab:vradohp11}, \ref{tab:vradohp12}, \ref{tab:vradspm12e}, and \ref{tab:vradspm12a}. The tables also give a mean RV for each observing run and the mean of the error intervals.}.

We note that the He\,\textsc{i}\,$\lambda\lambda$\,4143 and 4921 absorption lines have positive mean RVs, whilst the \he{i}{4471} absorption line has a negative mean RV. The \he{i}{4471} line also presents a strong variation compatible with the motion of a probable secondary component. For example, the mean RVs of \he{i}{4471} change from 36 (OHP 2011) to $-44$ (OHP 2012) and $-26$ (SPM 2012) \kms, while the mean RVs of \he{ii}{4542} change from 36 to 85 and 85 \kms, respectively (the error is of the order of $\pm10$ \kms).

It is also interesting to note that the \n{iv}{4058} emission line has a negative mean RV, whilst all the other emission lines have positive mean RVs. Several WN-type stars display such differences between  velocities of \n{iv}{4058} and other lines such as \he{ii}{4686} (see e.g. Shylaja \cite{Shylaja}). This can most probably be interpreted in terms of different formation sites, from close to the stellar surface to farther out in the wind.

\begin{figure}[ht!]
				\resizebox{\hsize}{!}{\includegraphics{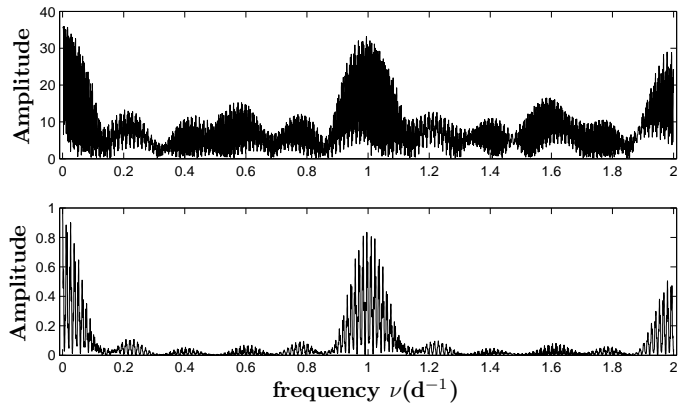}}
			\caption{\textit{Uppper panel}: Power spectrum computed on the basis of the RVs of the \n{v}{4604} line. \textit{Lower panel}: associated spectral window.}
			\label{fig:periodo4604}
		\end{figure}

We used our spectral time-series to search for the presence of a compact object in a close orbit. We first tried without success to fit a sine function to the series of RVs with the period of $2.3238$\,d proposed by Lamontagne et al. (\cite{Lamontagne}). We then applied the generalized Fourier technique developed by Heck et al. (\cite{Heck}), which was refined by Gosset et al. (\cite{Gosset2}), to the radial velocity time-series obtained from OHP and SPM data. The power spectrum did not present a significant peak in the frequency domain that would have been compatible with such a short period (see Fig.\,\ref{fig:periodo4604}). In conclusion, we did not find evidence for a short period in the WR\,138 system. We suggest that the line profile variations previously mentioned may be responsible for stochastic RV variations, which could have been mistaken for periodic changes.

		\begin{figure}[ht!]
				\resizebox{\hsize}{!}{\includegraphics{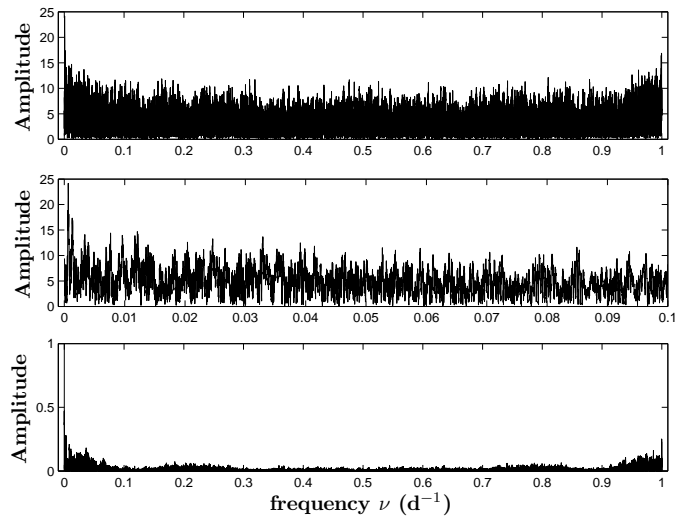}}
			\caption{\textit{Uppper and middle panels}: Power spectrum computed on the basis of the RVs of the \n{iv}{4058} line. The peak lies at $6.5\times10^{-4}$d$^{-1}$, which corresponds to a period of 1538.5 d. \textit{Lower panel}: associated spectral window (the highest value, 1, lies at $\nu=0$).}
			\label{fig:periodo}
		\end{figure}

Our data were obtained over a time basis that is too short to investigate time-scales as long as 1500 d. We therefore complemented our time-series with the measurements published by Massey (\cite{Massey}), Lamontagne et al. (\cite{Lamontagne}), and Annuk (\cite{Annuk}). We selected the spectral lines common to all studies. Even though this time-series is sparse and not homogeneous in terms of spectral resolution and signal-to-noise ratio, it offers the advantage of covering several decades. We caution, however, that the errors on the older measurements cannot be quantified, and neither do we have any information on the rest-wavelengths\footnote{The rest-wavelengths used in this paper come from the NIST Atomic Spectra Database available at: \url{http://www.nist.gov/pml/data/asd.cfm}.} used by previous authors to compute their RVs. Systematic biases are therefore possible. The results of the Fourier analysis of this extended time-series are interesting, however, because the power spectrum is clearly dominated by a peak at $6.5\times10^{-4}$d$^{-1}$ for the \n{iv}{4058} (see Fig.\,\ref{fig:periodo}) and \he{ii}{4542} lines. For the \he{ii}{4686} and \n{v}{4604} lines, the highest peaks are found at frequencies of $6.6\times10^{-4}$ and $6.7\times10^{-4}$d$^{-1}$. These peaks correspond to periods of 1538.5, 1515.2, and 1492.5 d. We estimate the uncertainty on the period as $\Delta P = 0.2\times\dfrac{P^2}{\Delta T}$, where $\Delta T$ is the time interval between the first and the last observations of the complete time-series (assuming the uncertainty on the peak frequency amounts to 20\% of the peak width), giving a value of $\sim35$\,d. In conclusion, we thus argue that WR\,138 is a binary system with a period of $1521.2\pm35$\,d, which agrees quite well with the value derived by Annuk (\cite{Annuk}, P = 1538 d).

	\begin{table}
		\caption{Radial velocities (in km\,s$^{-1}$) measured for the OHP 2011 campaign.}
		\label{tab:vradohp11}
		\centering
			\begin{tabular}{l c c c c}
			 \hline\hline
				HJD				 & He\,\textsc{i}					&  He\,\textsc{ii} 			 & N\,\textsc{v}				& S/N\\
				 - 2450000 & $\lambda$ 4471 & $\lambda$ 4542 & $\lambda$ 4604 &\\
				\hline
			5825.292& $49^{+6}_{-4}$ & $42^{+8}_{-6}$ & $36^{+4}_{-3}$ & 93\\
			5825.314& $34^{+3}_{-6}$ & $38^{+2}_{-2}$ & $40^{+4}_{-3}$ & 92\\
			5826.453& $29^{+2}_{-2}$ & $27^{+4}_{-3}$ & $42^{+4}_{-4}$ & 83\\
			5826.474& $3^{+4}_{-5}$ & $22.7^{+4}_{-2}$ & $32^{+7}_{-7}$ & 90\\
			5827.282& $11^{+2}_{-1}$ & $34.6^{+6}_{-9}$ & $51^{+8}_{-8}$ & 58\\
			5827.303& $72^{+16}_{-10}$ & $46.8^{+10}_{-13}$ & $54^{+9}_{-10}$ & 65\\
			5828.278& $30^{+6}_{-5}$ & $49^{+1}_{-1}$ & $35^{+6}_{-7}$ & 103\\
			5828.298& $47^{+5}_{-5}$ & $44.7^{+1}_{-2}$ & $35^{+9}_{-9}$ & 100\\
			5830.287& $45^{+4}_{-5}$ & $31.6^{+2}_{-3}$ & $39^{+8}_{-8}$ & 99\\
			5830.309& $44^{+5}_{-6}$ & $27^{+5}_{-5}$ & $46^{+3}_{-3}$ & 86\\
			\hline
			mean & $36^{+5}_{-5}$ & $36^{+4}_{-5}$ & $41^{+6}_{-6}$ & 87\\
			\hline
			\end{tabular}
	\end{table}

	\begin{table*}
		\caption{Same as Table \ref{tab:vradohp11}, but for the OHP 2012 campaign.}
		\label{tab:vradohp12}
		\centering
			\begin{tabular}{l c c c c c c}
			 \hline\hline
				HJD - 2450000 & \he{i}{4471} & \he{ii}{4542} & \n{v}{4604} & \he{ii}{4686} & H\,$\beta$ & S/N\\
				\hline
			6091.417 & $-37_{-1}^{+2}$ & $95_{-6}^{+7}$  & $94_{-2}^{+3}$  & $100_{-28}^{+12}$ & $90_{-14}^{+17}$ & 104\\
			6091.438 & $-57_{-9}^{+12}$ & $81_{-6}^{+3}$ & $98_{-4}^{+4}$  & $97_{-21}^{+17}$  & $73_{-22}^{+19}$ & 89\\
			6091.459 & $-30_{-1}^{+2}$ & $77_{-5}^{+3}$  & $91_{-4}^{+7}$  & $95_{-25}^{+17}$  & $69_{-9}^{+8}$ & 93\\
			6091.481 & $-51_{-7}^{+4}$ & $87_{-6}^{+7}$  & $88_{-4}^{+6}$  & $93_{-30}^{+16}$  & $71_{-15}^{+12}$ & 94\\
			6091.502 & $-35_{-6}^{+7}$ & $91_{-5}^{+3}$  & $91_{-4}^{+4}$  & $92_{-27}^{+17}$  & $70_{-11}^{+13}$ & 97\\
			6091.524 & $-49_{-1}^{+1}$ & $87_{-4}^{+4}$  & $87_{-4}^{+3}$  & $87_{-22}^{+19}$  & $62_{-10}^{+10}$ & 110\\
			6091.545 & $-41_{-3}^{+3}$ & $78_{-10}^{+9}$  & $83_{-8}^{+7}$  & $87_{-22}^{+18}$  & $61_{-17}^{+14}$ & 102\\
			6091.566 & $-46_{-4.5}^{+2}$ & $83_{-7}^{+7}$  & $87_{-12}^{+8}$ & $88_{-23}^{+17}$  & $58_{-22}^{+15}$ & 94\\
			6092.364 & $-41_{-15}^{+12}$ & $82_{-10}^{+16}$ & $93_{-2}^{+4}$  & $79_{-33}^{+19}$  & $43_{-13}^{+12}$ & 81\\
			6092.386 & $-26_{-3}^{+6}$ & $76_{-8}^{+10}$  & $99_{-5}^{+5}$  & $71_{-23}^{+25}$  & $37_{-10}^{+8}$ & 106\\
			6092.408 & $-33_{-1}^{+1}$ & $78_{-1}^{+3}$  & $100_{-7}^{+4}$  & $82_{-25}^{+17}$  & $52_{-7}^{+4}$ & 76\\
			6092.429 & $-26_{-1}^{+1}$ & $73_{-5}^{+6}$  & $88_{-8}^{+8}$  & $71_{-23}^{+23}$  & $47_{-9}^{+9}$ & 89\\
			6092.500 & / & $102_{-2}^{+12}$ & $86_{-1}^{+1}$ & $74_{-25}^{+22}$ & $61_{-8}^{+11}$ & 33\\
			6092.528 & / & $81_{-5}^{+5}$   & $92_{-5}^{+5}$   & $68_{-19}^{+16}$ & $36_{-10}^{+12}$ & 37\\
			6093.359 & $-62_{-7}^{+7}$ & $72_{-4}^{+2}$  & $95_{-10}^{+8}$  & $89_{-19}^{+15}$  & $57_{-7}^{+4}$ & 95\\
			6093.380 & $-40_{-1}^{+1}$ & $79_{-8}^{+3}$  & $93_{-4}^{+8}$  & $88_{-17}^{+15}$  & $55_{-18}^{+8}$ & 103\\
			6093.401 & $-42_{-4}^{+7}$ & $69_{-9}^{+6}$  & $91_{-4}^{+5}$  & $83_{-25}^{+20}$  & $61_{-25}^{+7}$ & 99\\
			6093.469 & $-51_{-2}^{+1}$ & $81_{-4}^{+4}$  & $98_{-4}^{+6}$  & $75_{-28}^{+23}$  & $56_{-19}^{+19}$ & 107\\
			6093.490 & $-41_{-5}^{+3}$ & $86_{-11}^{+5}$ & $101_{-6}^{+9}$ & $78_{-30}^{+21}$  & $69_{-13}^{+13}$ & 92\\
			6093.512 & $-44_{-5}^{+2}$ & $89_{-13}^{+8}$ & $100_{-5}^{+8}$  & $78_{-34}^{+22}$  & $61_{-26}^{+25}$ & 113\\
			6093.533 & $-45_{-4}^{+5}$ & $96_{-7}^{+4}$  & $100_{-5}^{+9}$  & $73_{-32}^{+26}$  & $64_{-28}^{+19}$ & 114\\
			6093.554 & $-40_{-3}^{+3}$ & $99_{-7}^{+8}$  & $96_{-6}^{+13}$ & $78_{-33}^{+22}$  & $61_{-24}^{+24}$ & 104.1\\
			6094.367 & $-32_{-4}^{+4}$ & $85_{-10}^{+4}$ & $97_{-4}^{+6}$  & $90_{-41}^{+22}$  & $78_{-16}^{+18}$ & 81\\
			6094.388 & $-38_{-3}^{+6}$ & $77_{-8}^{+6}$  & $101_{-7}^{+6}$ & $87_{-42}^{+23}$  & $77_{-14}^{+17}$ & 99\\
			6094.410 & $-58_{-1}^{+1}$ & $89_{-9}^{+4}$  & $101_{-8}^{+7}$ & $85_{-40}^{+25}$  & $75_{-21}^{+17}$ & 101\\
			6094.478 & $-46_{-2}^{+3}$ & $86_{-12}^{+7}$ & $100_{-3}^{+5}$  & $87_{-36}^{+21}$  & $75_{-14}^{+18}$ & 107\\
			6094.499 & $-43_{-3}^{+3}$ & $90_{-11}^{+8}$ & $99_{-4}^{+4}$  & $91_{-35}^{+19}$  & $79_{-13}^{+10}$ & 112\\
			6094.520 & $-53_{-5}^{+4}$ & $85_{-9}^{+6}$  & $96_{-4}^{+8}$  & $87_{-35}^{+23}$  & $77_{-8}^{+4}$ & 104\\
			6094.544 & $-43_{-9}^{+10}$ & $81_{-6}^{+8}$  & $4_{-3}^{+6}$  & $87_{-35}^{+24}$  & $64_{-6}^{+7}$ & 94\\
			6095.367 & $-56_{-2}^{+2}$ & $89_{-3}^{+7}$  & $108_{-3}^{+2}$ & $77_{-32}^{+25}$  & $78_{-11}^{+6}$ & 106\\
			6095.387 & $-42_{-2}^{+4}$ & $91_{-3}^{+2}$  & $100_{-4}^{+5}$ & $77_{-31}^{+26}$  & $74_{-9}^{+8}$ & 91\\
			6095.408 & $-45_{-4}^{+4}$ & $89_{-8}^{+12}$ & $97_{-3}^{+6}$  & $80_{-34}^{+23}$  & $67_{-9}^{+8}$ & 104\\
			6095.478 & $-42_{-8}^{+7}$ & $82_{-2}^{+5}$  & $98_{-3}^{+5}$  & $81_{-33}^{+24}$  & $47_{-15}^{+16}$ & 101\\
			6095.499 & $-28_{-3}^{+3}$ & $90_{-5}^{+7}$  & $97_{-5}^{+8}$  & $78_{-30}^{+26}$  & $47_{-10}^{+13}$ & 111\\
			6095.524 & $-54_{-3}^{+5}$ & $88_{-6}^{+8}$  & $92_{-5}^{+10}$ & $78_{-25}^{+25}$  & $54_{-12}^{+10}$ & 106\\
			6095.546 & $-51_{-5}^{+3}$ & $85_{-3}^{+2}$  & $91_{-4}^{+9}$  & $76_{-26}^{+28}$  & $60_{-12}^{+14}$ & 101\\
			6096.366 & $-47_{-10}^{+5}$ & $81_{-5}^{+8}$  & $96_{-7}^{+9}$  & $95_{-28}^{+17}$  & $79_{-14}^{+10}$ & 111\\
			6096.387 & $-47_{-10}^{+4}$ & $81_{-3}^{+2}$  & $96_{-5}^{+7}$  & $92_{-30}^{+20}$  & $78_{-17}^{+12}$ & 146\\
			6096.408 & $-31_{-8}^{+11}$ & $85_{-4}^{+7}$ & $88_{-5}^{+10}$ & $88_{-28}^{+20}$  & $76_{-16}^{+14}$ & 114\\
			6096.429 & $-53_{-6}^{+4}$ & $91_{-9}^{+10}$ & $106_{-2}^{+4}$ & $93_{-30}^{+15}$  & $78_{-19}^{+15}$ & 112\\
			6096.499 & $-40_{-1}^{+1}$ & $89_{-4}^{+5}$  & $89_{-11}^{+11}$ & $83_{-28}^{+22}$  & $78_{-18}^{+23}$ & 115\\
			6096.519 & $-55_{-34}^{+7}$ & $91_{-3}^{+3}$  & $95_{-3}^{+6}$  & $82_{-28}^{+23}$  & $88_{-23}^{+25}$ & 116\\
			6096.541 & $-59_{-8}^{+6}$ & $97_{-4}^{+5}$  & $100_{-3}^{+3}$  & $87_{-23}^{+17}$  & $91_{-25}^{+28}$ & 97\\
			\hline
			mean     & $-44_{-5}^{+4}$ & $85_{-6}^{+6}$ & $95_{-5}^{+6}$ & $84_{-29}^{+21}$ & $66_{-15}^{+13}$ & 99\\
		  	\hline
			\end{tabular}
	\end{table*}

	\begin{table*}
		\caption{Same as Table \ref{tab:vradohp11}, but for the SPM 2012 campaign (emission lines).}
		\label{tab:vradspm12e}
		\centering
			\begin{tabular}{l c c c c c c}
			 \hline\hline
				HJD - 2450000 & \n{iv}{4058} & \he{ii}{4100} & \he{ii}{4200} & \he{ii}{4542} & \he{ii}{4686} & H\,$\beta$\\
				\hline
				6167.717& $-85^{+12}_{-17}$ & $119^{+4}_{-3}$ & $144^{+0}_{-0}$ & $102^{+8.6}_{-6}$ & $107^{+0}_{-1}$ & $109^{+11}_{-6}$\\
				6167.732& $-104^{+16}_{-10}$ & $205^{+7}_{-8}$ & $145^{+1}_{-1}$ & $92^{+8}_{-7}$ & $98^{+0}_{-0}$ & $133^{+6}_{-6}$\\
				6167.747& $-79^{+18}_{-14}$ & $84^{+27}_{-17}$ & $141^{+2}_{-2}$ & $98^{+6}_{-5}$ & $108^{+1}_{-0}$ & $94^{+6}_{-5}$\\
				6168.658& $-70^{+4}_{-5}$ & $86^{+26}_{-17}$ & $147^{+2}_{-1}$ & $108^{+11}_{-11}$ & $113^{+0}_{-0}$ & $91^{+6}_{-6}$\\
				6168.673& $-98^{+6}_{-17}$ & $193^{+4}_{-6}$ & $155^{+1}_{-1}$ & $118^{+7}_{-5}$ & $113^{+1}_{-1}$ & $125^{+5}_{-3}$\\
				6168.689& $-85^{+11}_{-12}$ & $19^{+20}_{-12}$ & $142^{+1}_{-1}$ & $106^{+10}_{-16}$ & $111^{+1}_{-1}$ & $113^{+3}_{-3}$\\
				6168.745& $-93^{+16}_{-14}$ & $152^{+4}_{-3}$ & $152^{+0}_{-0}$ & $102^{+12}_{-10}$ & $112^{+1}_{-1}$ & $125^{+4}_{-4}$\\
				6168.756& $-86^{+15}_{-13}$ & $149^{+11}_{-15}$ & $154^{+1}_{-1}$ & $103^{+9}_{-8}$ & $113^{+2}_{-1}$ & $118^{+6}_{-4}$\\
				6168.771& $-90^{+11}_{-11}$ & $150^{+17}_{-18}$ & $149^{+2}_{-1}$ & $99^{+9}_{-10}$ & $114^{+0}_{-0}$ & $118^{+4}_{-5}$\\
				6169.640& $-79^{+17}_{-13}$ & $121^{+13}_{-11}$ & $138^{+3}_{-3}$ & $86^{+8}_{-8}$ & $109^{+2}_{-2}$ & $83^{+6}_{-6}$\\
				6169.665& $-90^{+19}_{-21}$ & $178^{+15}_{-14}$ & $135^{+0}_{-0}$ & $68^{+8}_{-9}$ & $112^{+5}_{-4}$ & $87^{+6}_{-5}$\\
				6169.682& $-92^{+23}_{-23}$ & $262^{+12}_{-7}$ & $119^{+2}_{-2}$ & $58^{+9}_{-7}$ & $101^{+3}_{-2}$ & $116^{+4}_{-4}$\\
				6170.722& $-82^{+6}_{-14}$ & $184^{+3}_{-3}$ & $137^{+3}_{4}$ & $72^{+7}_{-10}$ & $142^{+2}_{-2}$ & $117^{+7}_{-7}$\\
				6170.736& $-84^{+12}_{-22}$ & $222^{+7}_{-24}$ & $127^{+4}_{-6}$ &/& /&/\\
				6172.647& $-61^{+17}_{-14}$ & $32^{+2}_{-1}$ & $128^{+1}_{-1}$ & $100^{+6}_{-11}$ & $106^{+1}_{-1}$ & $111^{+8}_{-6}$\\
				6172.665& $-70^{+17}_{-19}$ & $92^{+3}_{-3}$ & $134^{+2}_{-3}$ & $98^{+6}_{-8}$ & $107^{+1}_{-1}$ & $107^{+8}_{-6}$\\
				6172.682& $-70^{+15}_{-9}$ & $50^{+2}_{-2}$ & $119^{+11}_{-24}$ & $90^{+4}_{-6}$ & $108^{+0}_{-0}$ & $104^{+4}_{-4}$\\
				6172.740& $-84^{+6}_{-8}$ & $83^{+3}_{-3}$ & $129^{+4}_{-4}$ & $93^{+7}_{-6}$ & $102^{+0}_{-0}$ & $97^{+5}_{-3}$\\
				6172.765& $-108^{+5}_{-8}$ & $57^{+2}_{-2}$ & $110^{+2}_{-2}$ & $70^{+5}_{-6}$ & $117^{+2}_{-1}$ & $122^{+6}_{-6}$\\
				6172.778& $-96^{+7}_{-13}$ & $87^{+5}_{-5}$ & $101^{+7}_{-15}$ & $96^{+4}_{-6}$ & $100^{+1}_{-1}$ & $105^{+8}_{-5}$\\
				6173.680& $-76^{+17}_{-12}$ & $27^{+2}_{-2}$ & $135^{+1}_{-1}$ & $101^{+7}_{-6}$ & $112^{+1}_{-1}$ & $106^{+4}_{-4}$\\
				6173.691& $-66^{+11}_{-9}$ & $-44^{+3}_{-3}$ & $112^{+2}_{-2}$ & $80^{+3}_{-2}$ & $119^{+2}_{-2}$ & $76^{+3}_{-2}$\\
				6173.706& $-66^{+10}_{-11}$ & $17^{+1}_{-1}$ & $101^{+4}_{-5}$ & $91^{+5}_{-7}$ & $110^{+0}_{-0}$ & $102^{+2}_{-2}$\\
				6173.725& $-91^{+12}_{-16}$ & $109^{+6}_{-4}$ & $124^{+1}_{-1}$ & $83^{+6}_{-6}$ & $108^{+1}_{-1}$ & $106^{+6}_{-5}$\\
				6173.820& $-88^{+6}_{-9}$ & $46^{+3}_{-3}$ & $116^{+1}_{-1}$ & $79^{+5}_{-5}$ & $102^{+2}_{-1}$ & $79^{+5}_{-5}$\\
				6173.831& $-99^{+19}_{-25}$ & $115^{+3}_{-3}$ & $124^{+1}_{-1}$ & $71^{+8}_{-10}$ & $99^{+1}_{-1}$ & $97^{+9}_{-6}$\\
				6173.845& $-101^{+4}_{-7}$ & $103^{+2}_{-2}$ & $119^{+2}_{-4}$ & $40^{+4}_{-6}$ & $100^{+1}_{-1}$ & $86^{+8}_{-7}$\\
			\hline
			mean & $-85^{+12}_{-14}$ & $112^{+8}_{-7}$ & $131^{+2}_{-3}$ & $85^{+7}_{-7}$ & $105^{+1}_{-1}$ & $101^{+5}_{-5}$\\
				\hline
			\end{tabular}
	\end{table*}
	
	\begin{table}
		\caption{Same as Table \ref{tab:vradohp11}, but for the SPM 2012 campaign (absorption lines).}
		\label{tab:vradspm12a}
		\centering
			\begin{tabular}{l c c c c}
			 \hline\hline
				HJD & He\,\textsc{i} &  He\,\textsc{i} & He\,\textsc{i} & S/N\\
				$-2450000$ & $\lambda$ 4143 &  $\lambda$ 4471 & $\lambda$ 4921 & \\
				\hline
			6167.717& $72^{+3}_{-4}$ 		& $-30^{+1}_{-1}$ & $75^{+7}_{-7}$ & 37\\
			6167.732& $116^{+39}_{-29}$ & $-44^{+4}_{-3}$ & $87^{+4}_{-4}$ & 38\\
			6167.747& $82^{+7}_{-6}$ 		& $-7^{+3}_{-3}$ & $36^{+1}_{-1}$ & 33\\
			6168.658& $115^{+15}_{-14}$ & $-6^{+8}_{-4}$ & $54^{+5}_{-4}$ & 35\\
			6168.673& $90^{+26}_{-24}$ 	& $-30^{+4}_{-4}$ & $132^{+4}_{-3}$ & 49\\
			6168.689& $82^{+3}_{-3}$ 		& $-33^{+4}_{-4}$ & $109^{+3}_{-3}$ & 38\\
			6168.745& $35^{+13}_{-12}$  & $-43^{+5}_{-3}$ & $71^{+7}_{-6}$ & 44\\
			6168.756& $92^{+6}_{-5}$   	& $-1^{+12}_{-9}$ & $85^{+5}_{-5}$ & 42\\
			6168.771& $111^{+2}_{-2}$   & $-21^{+4}_{-6}$ & $94^{+6}_{-7}$ & 49\\
			6169.640& $42^{+1}_{-1}$ 		& $-24^{+5}_{-5}$ & $80^{+8}_{-6}$ & 54\\
			6169.665& $71^{+9}_{-12}$ 	& $-22^{+5}_{-4}$ & $126^{+5}_{-7}$ & 47\\
			6169.682& $134^{+20}_{-17}$ & $-48^{+1}_{-0}$ & $117^{+8}_{-8}$ & 52\\
			6170.722& $41^{+3}_{-3}$ 		& $-3^{+11}_{-9}$ & $88^{+16}_{-13}$ & 52\\
			6170.736& $-149^{+27}_{-38}$ &/&/ & 45\\
			6172.647& $-13^{+5}_{-3}$ 	& $-12^{+3}_{-4}$ & $47^{+4}_{-3}$ & 41\\
			6172.665& $70^{+15}_{-12}$ 	& $-22^{+7}_{-4}$ & $69^{+4}_{-3}$ & 42\\
			6172.682& $-4^{+24}_{-18}$ 	& $-33^{+7}_{-5}$ & $48^{+4}_{-3}$ & 31\\
			6172.740& $-40^{+1}_{-1}$ 	& $-42^{+4}_{-4}$ & $39^{+6}_{-7}$ & 37\\
			6172.765& $-120^{+1}_{-1}$ 	& $-20^{+1}_{-1}$ & $91^{+1}_{-1}$ & 30\\
			6172.778& $-46^{+31}_{-22}$ & $-4^{+3}_{-3}$ & $55^{+4}_{-2}$ & 42\\
			6173.680& $54^{+6}_{-5}$ 		& $-31^{+2}_{-1}$ & $36^{+4}_{-8}$ & 28\\
			6173.691& $18^{+2}_{-1}$ 		& $-36^{+3}_{-2}$ & $-55^{+3}_{-3}$ & 21\\
			6173.706& $162^{+0}_{-0}$		& $-69^{+1}_{-1}$ & $20^{+2}_{-2}$ & 31\\
			6173.725& $39^{+9}_{-9}$ 		& $-40^{+4}_{-3}$ & $72^{+4}_{-3}$ & 40\\
			6173.820& $46^{+1}_{-1}$	  & $-14^{+7}_{-5}$ & $34^{+3}_{-3}$ & 41\\
			6173.831& $37^{+7}_{-7}$ 		& $-29^{+3}_{-3}$ & $61^{+8}_{-9}$ & 46\\
			6173.845& $84^{+10}_{-7}$	  & $-34^{+4}_{-3}$ & $76^{+2}_{-3}$ & 54\\
			\hline
			mean & $45^{+11}_{-9}$ & $-26^{+4}_{-4}$ & $65^{+5}_{-5}$ & 41\\
			\hline
			\end{tabular}
	\end{table}

\section{X-ray emission}
	
	\subsection{Spectral fits}
To determine the nature of the system, we fitted several models to the EPIC spectra. Very few individual lines are visible, which is due to the low resolution. The fits were calculated with the XSPEC software, v 12.6.0 (Arnaud \cite{Arnaud}). Our spectra present the highest emission peak around $0.8-1$\,keV, but extend to more or less 8\,keV, which is quite high assuming a (single) O-type star, but is more typical of a colliding-wind binary. The X-ray emission of an individual massive star can be, in first approximation, described by an optically thin thermal plasma heated to a few $10^6$K by shocks intrinsic to its stellar wind (Feldmeier et al. \cite{Feldmeier}, \& White \cite{LucyWhite}, Lucy \cite{Lucy}, Owocki et al. \cite{Owocki}). Whether or not the same mechanism also operates in the winds of individual Wolf-Rayet stars is currently still an open question. In fact, unlike O-type stars, single WRs do not exhibit a clear relation between their X-ray luminosity and their stellar wind parameters (Wessolowski \cite{Wessolowski}). While some (apparently) single WR stars are detected as X-ray sources, others remain undetected, even with \textit{XMM-Newton} and \textit{Chandra} (e.g. Oskinova \cite{Oskinova03}, Gosset et al. \cite{Gosset05}, Skinner et al. \cite{Skinner10,Skinner12}). In the case of a WR+OB binary system, an additional site for thermal X-ray production is the colliding-wind region, where strong shocks can heat the plasma to temperatures of several $10^7$K (Pittard \& Parkin \cite{Pittard}). In addition, a power-law component may be required to account for a possible inverse-Compton-scattering emission that can occur in the wind-collision region (De Becker \cite{DeBecker}). Finally, a combination of an optically thin thermal plasma and a black-body emission can be considered in the case of a WR + neutron star system.
We investigated all these models, restricting our analyses to the energy range 0.3-10\,keV to avoid data with very low S/N ratio. In the third {\it XMM-Newton} observation, WR\,138 partially fell in a gap of the pn detector. We tried to fit the data, but the results obtained with this set give a very poor quality fit. To avoid a misinterpretation, we therefore decided to reject these pn data. We adopted an equivalent interstellar column of neutral hydrogen of $N_{H,ism} = 0.58\times 10^{22}$ E(B-V)$= 0.37 \times 10^{22}$cm$^{-2}$ (Bohlin et al. \cite{Bohlin}) using the star's colour excess obtained from E(B-V) = 1.21/4.1 A$_{\text{v}}$ = 0.63 mag (van der Hucht \cite{Hucht}). WR stars are evolved objects and are characterized by non-solar abundances. We therefore used models with adjustable abundances. However, in view of the low spectral resolution, we had to reduce the number of free parameters to avoid local minima or non-physical solutions. We therefore fixed the abundances to common values for WN stars found in the literature (Crowther, Smith \& Hillier \cite{Crowther}, Hamann \& Koesterke \cite{HamannK}, Hamann \& Gr\"{a}fener \cite{HamannG}). These cases correspond to a H/He ratio equal to 0.05, 0.15, and 0.25. The C/He, N/He, and O/He are equal to 0.00048, 0.048, and 0, while all other elements were kept fixed to their solar value. Table \ref{tab:abundances} recalls the relative abundances adopted for the different chemical elements.	
	
	\begin{table}
		\caption{Relative and solar abundances used for the X-ray spectral modelling.}
		\label{tab:abundances}
		\centering
			\begin{tabular}{l c c c}
				\hline\hline
				 & H/He = 0.25 & H/He = 0.15 & H/He = 0.05\\
				 \hline
				(He/H)/(He/H)$_{\sun}$ & $46.998$ & $78.330$ & $234.990$ \\
				(N/H)/(N/H)$_{\sun}$   & $23.077$ & $38.462$ & $115.385$ \\
				(C/H)/(C/H)$_{\sun}$   & $5.818$ & $9.697$ & $29.091$ \\
				(O/H)/(O/H)$_{\sun}$   & $0.000$ & $0.000$ & $0.000$ \\
				(X/H)/(X/H)$_{\sun}$   & $4.000$ & $6.666$ & $20.000$ \\
		  	\hline
		  	\multicolumn{4}{l}{Solar abundances (Grevesse \& Sauval \cite{Grevesse})}\\
		  	\hline
		  	(He/H)$_{\sun}$ & 0.08511 & & \\
		  	(N/H)$_{\sun}$  & 0.00832 & & \\
		  	(C/H)$_{\sun}$  & 0.00033& & \\
				(O/H)$_{\sun}$  & 0.00068& & \\
		  	\hline
			\end{tabular}
			\tablefoot{X stand for the other metal elements (Fe, Ni, Ar, etc).}
	\end{table}
	
We first tried a single-temperature plasma model (APEC) and an intrinsic absorption column. As expected, this simple model gives poor quality fits, and therefore we switched to a more complex two-temperature model with an intrinsic absorption column (wabs$_{ISM}$*wabs*(APEC+APEC)). We point out that the absorption is dominated by the interstellar absorption, therefore we used a wabs model with solar abundances (Anders \& Ebihara \cite{Anders}) instead of a vphabs model with non-solar abundances. The two models led to solutions with similar reduced $\chi^2$, temperatures, normalization factors, and a low $N_H$. The three sets of non-solar abundances yielded solutions with similar reduced $\chi^2$, temperatures, and intrinsic absorption column. The normalization factors were different, however, which is expected because in modifying the relative abundances, we changed the number of available free electrons and hence the emission measure of the plasma. We therefore adopted the intermediate value of the abundance for the rest of the study (i.e. H/He = 0.15). Table \ref{tab:2TAPEC} gives a summary of the best-fit parameters. If we simultaneously fit the data from the two detectors (except for the third observation where the pn data were rejected), the results of the fits suggest a weaker variability than in the case of the fits of the MOS\,2 data alone.

		\begin{figure}[ht!]
		  \resizebox{\hsize}{!}{\includegraphics{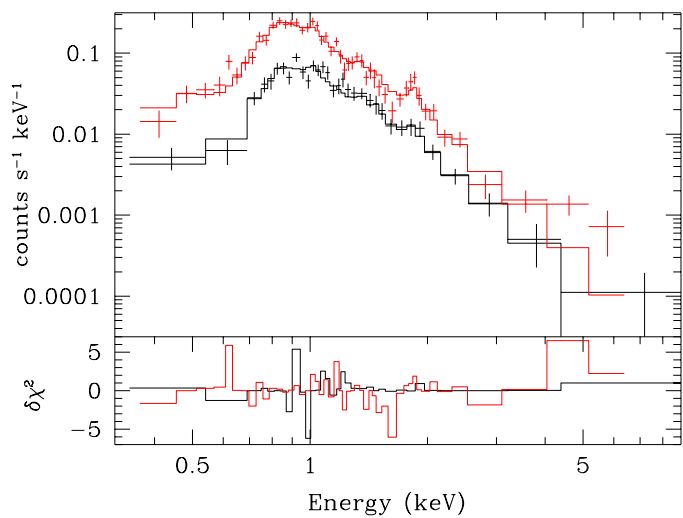}}
			\caption{\textit{Top panel}: Spectra (MOS2, \textit{in black}, and pn, \textit{in red}, instrument) of the {\it XMM-Newton} observation of HJD=2455694.768 and best-parameter fit obtained with {\it XSPEC}. \textit{Lower panel}: $\chi^2$ increments multiplied by the sign of the data -- model difference.}
			\label{fig:2}
		\end{figure}

We also used a one-temperature plasma plus a power-law model with an intrinsic absorption column (wabs$_{ISM}$*wabs*(APEC+power-law)). The quality of the fit is similar to those obtained with the previous model except for the third observation, for which this model gives a better fit. Notice that the poorer response at energy $<0.5$ and $>5$ keV mainly explains the high $\chi^2$ of the 2-T plasma and the 1-T plasma plus power-law model fit. Table \ref{tab:APECPL} gives a summary of the best-fit parameters. Because of the low S/N ratio above 2 keV, it is difficult to determine which of these models is best-suited to the WR\,138 system. It has to be stressed that the photon index and normalization of the power-law component are found to be highly variable in this model. In particular, we had to deal with multiple local minima that severely affected our capability to derive consistent parameters from one data set to the other. It is clear, for instance, that the fit in some cases predicted excessive normalization parameter values, which were therefore accompanied by large absorption columns to compensate the flux excess at lower energies. This probably points to a significant inadequacy of the 1T+power-law model. Finally, the APEC plus black-body model gives low-quality fits ($\chi_{\nu}^2\geq 2$). 

	\begin{table*}[ht]
		\caption{Best-fit parameter results for the two-temperature plasma model.}
		\label{tab:2TAPEC}
		\centering
			\begin{tabular}{l c c c c c c}
				\hline\hline
				HJD & $n_H$ & $kT_1$ & Norm$_1$ & $kT_2$ & Norm$_2$ & $\chi_{\nu}^2$ (d.o.f.) \\
				-2440000 & ($10^{20}$cm$^{-2}$) & (keV) & ($10^{-5}$cm$^{-5}$) & (keV) & ($10^{-5}$cm$^{-5}$)\\
				\hline\\[-3mm]
				{\it XMM-Newton}\\
				15686.713 & $5.74^{+3.53}_{-3.75}$ & $0.64\pm0.04$ & $5.98^{+5.69}_{-1.24}$ & $1.27^{+0.36}_{-0.15}$ & $0.72^{+0.57}_{-0.09}$&$1.40$ (88)\\ [3mm]
				15694.768 & $< 3.98$ & $0.63\pm0.04$ & $4.15^{+1.17}_{-0.42}$ & $1.27^{+0.21}_{-0.13}$ & $1.81^{+0.47}_{-0.47}$ & $1.13$ (78)\\[3mm]
				15700.152 & $< 2.35$ & $0.65^{+0.08}_{-0.05}$ & $4.41^{+1.84}_{-0.84}$ & $1.42^{+0.38}_{-0.29}$ & $1.99^{+0.94}_{-0.98}$ & $1.89$ (32)\\[1mm]
		  	\hline
				{\it CHANDRA}\\
				15213.113 & $< 1.1$ & $0.58^{+0.03}_{-0.07}$ & $5.4^{+0.44}_{-0.82}$ & $1.12^{+0.11}_{-0.13}$& $1.94^{+8.75}_{-0.38}$ & $1.20$ (170)\\[1mm]
				15213.113 & $< 0.8$ & $0.64$ (fixed) & $5.8^{+0.30}_{-0.32}$  & $1.32$ (fixed) & $1.28^{+0.23}_{-0.21}$ & $1.27$ (172)\\[1mm]
				\hline
				{\it ROSAT}\\
				6928.585 & $< 13.0$ & $0.64$ (fixed) & $6.91^{+3.43}_{-1.48}$ & $1.32$ (fixed) & $1.5$ (fixed) & $1.42$ (17)\\[1mm]		
				9508.604 & $< 15.7$ & $0.64$ (fixed) & $5.21^{+3.06}_{-1.42}$ & $1.32$ (fixed) & $1.5$ (fixed) & $1.12$ (17)\\[1mm]	
				\hline
			\end{tabular}
			\tablefoot{The best-fit parameters of the {\it XMM-Newton} observation correspond to a simultaneous fit of the MOS2 and pn spectra (except for the last observation where the pn data were rejected). The errors correspond to the 90\% confidence interval. All fits were performed including an ISM absorption column of $37\times 10^{20}$cm$^{-2}$.}
	\end{table*}
	
	\begin{table*}[ht]
		\caption{Same as Tab. \ref{tab:2TAPEC}, but for the one-temperature plasma plus power-law model.}
		\label{tab:APECPL}
		\centering
			\begin{tabular}{l c c c c c c}
				\hline\hline
				HJD & $n_H$ & $kT_1$ & Norm$_1$ & PhoIndex & Norm$_2$ & $\chi_{\nu}^2$ (d.o.f.)\\
				-2440000 & ($10^{20}$cm$^{-2}$) & (keV) & ($10^{-5}$cm$^{-5}$) &  & ($10^{-5}$ph keV$^{-1}$cm$^{-2}$s$^{-1}$) & \\
				\hline\\[-3mm]
				{\it XMM-Newton}\\
				15686.713 & $16.6^{+4.25}_{-4.36}$ & $0.64\pm0.03$ & $10.1^{+1.59}_{-1.74}$ & $1.16^{+0.51}_{-0.96}$ & $1.32^{+3.56}_{-1.04}$&$1.39$ (88)\\ [3mm]
				15694.768 & $8.21^{+5.28}_{-4.27}$ & $0.64\pm0.03$ & $6.24^{+2.97}_{-1.23}$ & $2.67^{+0.71}_{-0.57}$ & $8.03^{+6.30}_{-7.45}$&$1.29$ (78)\\[3mm]
				15700.152 & $13.3^{+9.94}_{-7.92}$&$0.63^{+0.08}_{-0.06}$&$6.49^{+3.03}_{-1.86}$&$3.47^{+1.26}_{-1.39}$&$17.7^{+14.6}_{-14.7}$&$0.99$ (32)\\[1mm]
		  	\hline
				{\it CHANDRA}\\
				15213.113 & $2.26^{+4.99}_{-2.26}$ & $0.64$ (fixed) & $5.45^{+0.93}_{-0.65}$ & $3.93^{+0.72}_{-0.44}$ & $ 2.26^{+4.99}_{-2.26} $ & $1.28$ (171)\\[1mm]
				\hline
				{\it ROSAT}\\
				6928.585& $< 28.8$ & $0.64$ (fixed) & $7.31\pm6.41$ & $2.43$ (fixed) & $9.2$ (fixed) & $1.42$ (17)\\[1mm]	
				9508.604 & $< 32.9$ & $0.64$ (fixed) & $5.64\pm5.64$ & $2.43$ (fixed) & $9.2$ (fixed) & $1.25$ (17)\\[1mm]		
				\hline
			\end{tabular}
			\tablefoot{Same note as Table \ref{tab:2TAPEC}}
	\end{table*}
		
In conclusion, the two-temperature plasma model seems to be the most adequate one. This model is dominated by the lowest temperature at 0.6\,keV. Hamann et al. (\cite{Hamann95}) reported a luminosity of $\log\text{L}_{bol} = 38.9$ (assuming a photometric distance of $\text{d}=1.82$\,kpc, Lundstr\"{o}m \& Stenholm \cite{Lundstrom}) for WR\,138, which is well inside the range of values of the other WN5-6 stars, that is, $\log\text{L}_{bol} =[38.5, 39.2]$. The X-ray luminosity that we derive with the distance used by Hamann et al. (\cite{Hamann95}) is about $\log\text{L}_x=32.65$. With this value, we can derive a $\log(\text{L}_x/\text{L}_{bol})$ ratio of $-6.25$. Oskinova (\cite{Oskinova}) suggested a distance of $\text{d}=1.26$\,kpc and a $\log\text{L}_{bol}=38.88$. With this distance and bolometric luminosity, the X-ray luminosity lies in the range $\log\text{L}_x=[32.33,32.48]$, which leads to a $\log(\text{L}_x/\text{L}_{bol})=[-6.55,-6.41]$.

The X-ray analysis shows that WR\,138 is no exceptional star in this wavelength domain. Oskinova (\cite{Oskinova}) reported on other WN+OB star binaries whose spectra could be represented by a two-temperature plasma model dominated by a soft component $kT_1\approx 0.6$\,keV and a hard component $kT_2\approx 2$\,keV. Our modelling agrees well for the temperature of the soft component, but gives a sightly lower value for the temperature of the hard component.

Our results (moderate X-ray luminosity, X-ray spectral morphology) argue against the presence of an accreting compact companion. They instead suggest a more conventional long-period WR + OB binary.

	\begin{table*}[ht]
		\caption{X-ray flux.}
		\label{tab:flux}
		\centering
			\begin{tabular}{l c c c c c c c c c}
				\hline\hline\\[-3mm] 
				HJD & \multicolumn{4}{c}{2-T plasma model ($10^{-13}$erg\,s$^{-1}$\,cm$^{-2}$)} & & \multicolumn{4}{c}{1-T plasma +power-law model ($10^{-13}$erg\,s$^{-1}$\,cm$^{-2}$)}\\
						 \cline{2-5} \cline{7-10}\\[-3mm] 
				-2440000 & F$^{\text{corr}}_{\text{x,s}}$ & F$^{\text{corr}}_{\text{x,m}}$ & F$^{\text{corr}}_{\text{x,h}}$ & F$^{\text{obs}}_{\text{x,tot}}$ & & F$^{\text{corr}}_{\text{x,s}}$ & F$^{\text{corr}}_{\text{x,m}}$ & F$^{\text{corr}}_{\text{x,h}}$ & F$^{\text{obs}}_{\text{x,tot}}$ \\[1mm] 
				\hline
				{\it XMM-Newton}\\
				15686.710 & 7.16 & 4.42 & 1.15 & $5.49\pm0.18$ & & 6.87 & 4.33 & 1.97 & $5.60\pm0.18$ \\
				15694.768 & 6.63 & 3.70 & 0.98 & $4.67\pm0.17$ & & 6.58 & 3.54 & 1.22 & $4.91\pm0.18$ \\
				15700.152 & 7.06 & 4.04 & 1.28 & $5.27\pm0.16$ & & 6.58 & 3.78 & 1.04 & $4.83\pm 0.16$ \\
				\hline
				{\it CHANDRA}\\
				15213.113$^1$ & 8.51 & 4.24 & 0.87 & $5.30\pm0.1$ & &  &  &  &  \\
				15213.113$^2$ & 8.40 & 4.10 & 0.95 & $5.26\pm0.1$ & & 10.77 & 3.99 & 0.81 & $5.34\pm 0.1$ \\
				\hline
				{\it ROSAT}\\
				6928.585$^2$ & 9.97 & 4.91 &  & $5.28\pm0.58$ & & 10.73 & 4.70 & & $5.32\pm0.59$ \\
				9508.604$^2$ & 7.75 & 4.02 &  & $4.23\pm0.64$ & & 8.54 & 3.83 & & $4.29\pm0.64$ \\
				\hline
			\end{tabular}
			\tablefoot{F$^{\text{corr}}_{\text{x,s}}$: X-ray flux corrected for ISM$_{\text{abs}}$ in the band 0.5-1.0\,keV, F$^{\text{corr}}_{\text{x,m}}$ in the band 1.0-2.0\,keV, and F$^{\text{corr}}_{\text{x,h}}$ in the band 2.0-10.0\,keV. F$^{\text{obs}}_{\text{x,tot}}$:observed X-ray flux in the band 0.5-10.0\,keV (or 0.5-2.0\,keV for the {\it ROSAT} mission). $^1$Corresponds to the spectral fit where the temperatures are free parameters. $^2$Corresponds to the spectral fit with fixed temperatures.}
	\end{table*}
	
	\subsection{Long-term behaviour}

Several pointed observations from {\it SWIFT}, {\it ROSAT}, and {\it CHANDRA} (see Sect.\ref{sect2.1}) were also used to determine whether WR\,138 presents long-term variations in its X-ray emission. Such long-term variations exist in long-period highly eccentric WN + OB colliding wind binaries (e.g. WR\,22, Gosset et al. \cite{GossetWR22}, WR\,25, Pollock \& Corcoran \cite{PollockWR25}). We also fitted the two-temperature plasma model and the one-temperature plus power-law model to the {\it ROSAT} and {\it CHANDRA} data. For the {\it ROSAT} data, the spectra have a very low resolution and do not cover the energies beyond 2.2 keV. A fit of the data without fixing the temperature gives unrealistically high temperatures (kT$>10$keV). Therefore, we fixed the temperature to the mean value of the {\it XMM-Newton} best-fit temperatures. For the {\it CHANDRA} data, we tried to fit the data with or without fixed temperatures. In the first case (fixed temperatures), the normalization factors (and intrinsic absorption) are similar to the {\it XMM-Newton} fits. In the second case (no fixed temperature), the fit indicates slightly lower temperatures. These slight differences could be due to remaining cross-calibration uncertainties. The best-fit parameters are given in Tables \ref{tab:2TAPEC} and \ref{tab:APECPL}.

\begin{figure}[ht!]
		  \resizebox{\hsize}{!}{\includegraphics{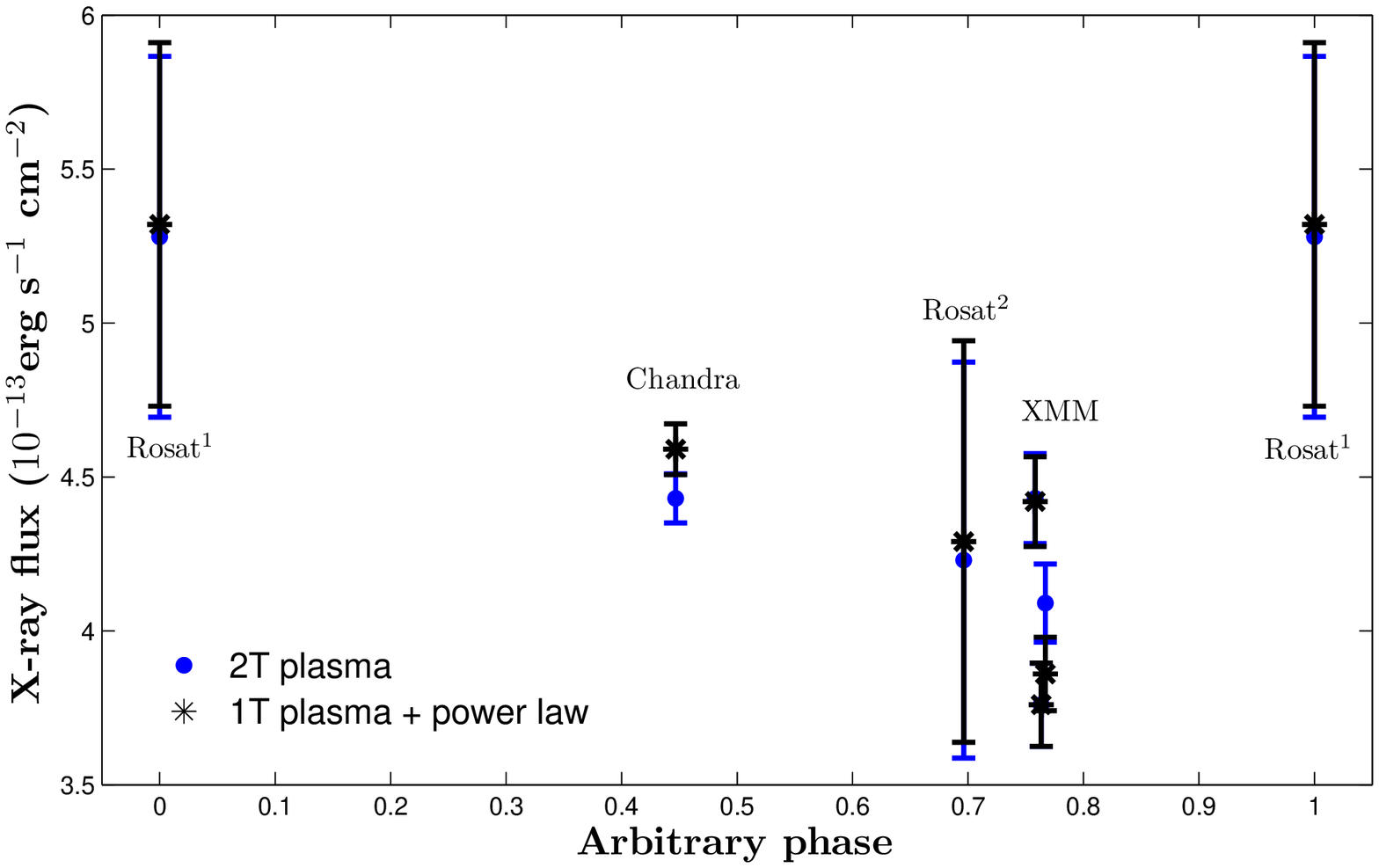}}
			\caption{Variation of the observed X-ray fluxes derived with {\it XSPEC} for the {\it ROSAT}, {\it CHANDRA}, and {\it XMM-Newton} observations in the energy band 0.5-2 keV. $^1${\it ROSAT} observation 2446928.585, $^2${\it ROSAT} observation 2449508.604. The phase 0 has been arbitrary chosen to the oldest {\it ROSAT} observation. We did not use the T$_0$ derived by Annuk (\cite{Annuk}) because they strongly depend on the line considered. The period has been fixed to P=1521.1 d. Error bars are related to the relative error of the count rate (see text).}
			\label{fig:flux}
		\end{figure}

From these fits, we computed the X-ray fluxes in three spectral bands: soft (0.5-1.0 keV), medium (1.0-2.0 keV), and hard (2.0-10.0 kev). These fluxes are quoted in Table \ref{tab:flux}. We assumed that the relative error on the observed X-ray fluxes is the same as for the count rates, and chose the oldest {\it ROSAT} observation as the starting point for arbitrary ephemeris (using the P=1521.1 d period found in Sect. 3). The error estimate is probably underestimated because some parameters such $N_{H,ism}$ are poorly known. The flux also strongly depends on the model used, which is affected by uncertainties. These additional errors are unfortunately difficult to estimate. We tried to use the \textit{flux err} command of XSPEC to calculate another evaluation of the error. The results are of the same order of magnitude (few percent) for the upper limit. The lower limit derived from XSPEC is much lower, but a model computed from
this lower limit cannot account for the observed flux. Fig.~\ref{fig:flux} shows that WR\,138 displays only modest, if any, flux variations, although because of the uncertainties on the long-term orbital solution, it is not clear whether or not our data sample critical phases (such as the periastron passage) of the orbital motion in the WN5-6 + OB binary. A dedicated monitoring of WR\,138 around periastron might reveal a significant increase in flux. However, this requires an accurate ephemeris and needs a dedicated and long-term optical monitoring of WR\,138.

Table \ref{tab:count} gives the value of the count rates given by XSPEC for the different missions (observed count rates) and those simulated with XSPEC using the model parameters of a simultaneous fit of all {\it XMM-Newton} spectra with a two-temperature plasma model (simulated count rates) folded through the appropriate instrumental response matrices. We stress that the count rates are those within the extraction regions and are therefore depending on where the source lies on the detector and the size of chosen regions; consequently, they are linked to the ancillary response files (arf). Because the exposure time and arf are different for each observation, we see small differences in the simulated count rates of the same instruments. The comparison between observed and simulated count rate is aimed at providing an estimate of the potential long-term variations between the epoch of the {\it XMM-Newton} observations and that of all previous observations. The count rates of the three {\it XMM-Newton} observations present variations of less than 10\%, that is, less than 3-$\sigma$ except for the first pn observation (see Table \ref{tab:count}). The first MOS2 observation also presents a quite high variation that is just below 3-$\sigma$. Therefore there is no contradiction between the pn and MOS2 results. The oldest {\it ROSAT} PSPC and HRI observations also display count-rate variations stronger than 3-$\sigma$, which are therefore considered to be significant. 

	\begin{table*}[ht]
		\caption{X-ray count rates.}
		\label{tab:count}
		\centering
			\begin{tabular}{l c c c c c c}
				\hline\hline
				Mission - Instrument & HJD & Observed count rate & Relative & Simulated count rate & Relative & Energy \\
														 &		 &										 & error 		&											 & deviation$^1$ & range \\
														 & -2440000 & ($10^{-2}$cts s$^{-1}$) & (\%) & ($10^{-2}$cts s$^{-1}$) & (\%) & keV\\
				\hline
				{\it XMM-Newton} MOS2 			& 15686.713 & $5.42\pm0.18$ & 3.3 & $4.97\pm0.15$ & 8.4 &  0.2-10 \\
																		& 15694.768 & $5.03\pm0.18$ & 3.6 & $5.00\pm0.16$ & 0.5 & 0.2-10 \\
																		& 15700.152 & $5.54\pm0.17$ & 3.1 & $5.14\pm0.15$ & 7.1 & 0.2-10 \\
				{\it XMM-Newton} pn					& 15686.713 & $17.95\pm0.41$ & 2.3 & $15.87\pm0.34$ & 11.6 & 0.2-10 \\
																		& 15694.768 & $15.28\pm0.40$ & 2.6 & $15.84\pm0.35$ & -3.7 & 0.2-10 \\
				\hline
				{\it EINSTEIN}-IPC$^2$			& 3972.050  & $1.00\pm0.40$ & 40.0 & $2.09\pm0.20$ & -109.4 & 0.2-4.5 \\
				{\it ROSAT} All Sky Survey 	& 8210.000  & $3.26\pm0.87^3$ & 26.7 & $3.20\pm0.57$ & 1.8 & 0.1-2 \\
				{\it ROSAT} PSPC 					  & 6928.585  & $4.50\pm0.50$ & 11.1 & $2.66\pm0.25$ & 40.8 & 0.1-2 \\
																		& 9508.604  & $3.30\pm0.50$ & 15.2 & $2.54\pm0.26$ & 23.1 & 0.1-2 \\
				{\it ROSAT} HRI 						& 9669.924  & $1.17\pm0.11$ & 9.8 & $1.68\pm0.06$ & -43.6 & 0.1-2 \\
				{\it SWIFT} 								& 15635.350 & $1.75\pm0.42$ & 24.1 & $1.20\pm0.24$ & 31.5 & 0.3-10 \\
				{\it CHANDRA} 							& 15213.113 & $6.58\pm0.12$ & 1.8 & $6.30\pm0.10$ & 4.3 & 0.2-10 \\
		  	\hline
			\end{tabular}
			\tablefoot{$^1$ The relative error corresponds to $100\times(\text{obs}-\text{simulation})/\text{obs}$. The count rates have been obtained with XSPEC. $^2$ Values from Pollock (\cite{Pollock}). $^3$ This value is in good agreement with the one derived by Pollock et al.\ (\cite{RASS}) who reported a {\it ROSAT} All Sky Survey PSPC count rate of $(3.26 \pm 0.87) \times 10^{-2}$ cts\,s$^{-1}$. Note that ROSAT HRI had little or no energy resolution.}
	\end{table*}

\section{Summary and conclusion}

Throughout this paper we have investigated the validity of a scenario where the WR\,138 system harbours a close compact companion. We found that neither the optical spectra nor the X-ray spectra analysis provided clear clues for such a companion.

The optical campaigns have shown that, as for many Wolf-Rayet stars, the broad emission lines of WR\,138 display variable subpeaks (L\'epine et al.\ \cite{Lepine}, see fig. \ref{fig:linesvar}) that might affect the determination of the radial velocity of the WN star. The Fourier analyses of RVs do not show any clear short period compatible with that derived by Lamontagne et al. (\cite{Lamontagne}). Photometric studies have also been conducted by Martin \& Plummer (\cite{MP}), Gaposchkin (\cite{Gap}), and Ross (\cite{Ross}), who first reported on the low-level irregular variability of WR\,138, and a photometric period of 11.6 d was reported by Moffat \& Shara (\cite{MS}). However, these results are based on limited and sparse time-series that are not adequate for distinguishing between periodic and sporadic variations such as are frequently found in photometric observations of WR stars (e.g.\ Gosset et al.\ \cite{Gosset}). Indeed, from the longer-term {\it Hipparcos} photometry of WR\,138, Marchenko et al.\ (\cite{Marchenko}) reported on stochastic variability. The X-ray analysis seem to point in the same direction, that is, the absence of a compact companion. Indeed, the luminosity of WR\,138 lies in the range $\log\text{L}_x=[32.33,32.48]$, while the luminosity of the Cyg X-3 system ($\log\text{L}_x\sim10^{38}$ergs$^{-1}$, Skinner et al. \cite{Skinner10}), which is the best-known system including a compact companion that accretes the wind material from a WR star, is several orders of magnitude higher.

The question of a long-period OB companion has also been raised. The presence of this OB companion is now ascertained and supported by both optical and X-ray analyses. The Fourier analysis of the new RVs added to the previously reported ones of Massey (\cite{Massey}), Lamontagne (\cite{Lamontagne}), and Annuk (\cite{Annuk}) clearly reveals a peak associated to a long period. Moreover, the variations of the measured RVs cannot be completely explained by intrinsic wind inhomogeneities. The most probable scenario is that WR\,138 is composed of a WN\,5-6 star with an OB-type companion in a wide eccentric orbit of about 1521 d. 

The X-ray spectra present an extension to high energies that would be unusual for single-star emission, but which is more common for a long-period system. The typical X-ray luminosity of WN-type stars displays large scatter. For example, WR\,25 displays an X-ray luminosity of $\log\text{L}_x=33.9$, while some other stars remain below detection level ($\log$L$_x<30.0$, Oskinova \cite{Oskinova}), such as WR\,40 ($\log\text{L}_x<31.6$, Gosset et al. \cite{Gosset05}). Oskinova (\cite{Oskinova}) reported that the typical value of $\log(\text{L}_x/\text{L}_{bol}) \text{ is} \sim-7$. Our results ($\log$L$_x=[32.33,32.48]$ and $\log(\text{L}_x/\text{L}_{bol})=[-6.55,-6.41]$) allow us to conclude that WR\,138 is not particularly luminous in X-rays, but it is not particularly faint either. In fact, WR\,138 seems to lie in the mean range of what we can observe for WN-type stars. This moderate X-ray luminosity also agrees well with a companion in a wide orbit. Skinner et al. (\cite{Skinner10,Skinner12}) and Oskinova (\cite{Oskinova}) also suggested that a two-temperature plasma model reproduces the spectra of a WN+OB binary system well. Our best-fit parameter again agrees well with this suggestion. 

We also noticed variation in the X-ray emission, but, because of the small number of observations, we cannot deduce a phase-locked variability, but we cannot exclude that the variation is due to a weak wind-wind interaction between the WN\,5-6 and an OB companion. Finally, the radio emission of WR\,138 was reported to be thermal by Montes et al.\ (\cite{Montes}) with a corresponding mass-loss rate of $1.18\,10^{-5}$\,M$_{\odot}$\,yr$^{-1}$ (assuming a distance of 1.4\,kpc). Montes et al.\ (\cite{Montes}) inferred a radio spectral index close to unity, higher than expected for the free-free emission of the wind of a single star. These authors suggested that this might reflect either a clumpy wind or be due to a radiative-wind interaction zone in a binary system, which agrees well with the scenario of a WN+OB companion.

In conclusion, we can confidently reject the scenario of a compact companion in a close orbit around the WN star because no clues (level of X-ray emission, lack of short-term RV variations) support this assumption. The OB companion proposed by Annuk (\cite{Annuk}) can be confirmed by both optical analysis (RV variations, peak in Fourier analysis) and X-ray analysis (moderate luminosity, extension of the spectra to high energies, variations). Finally, the orbital period can be refined to $\text{P}=1521.2\pm35$\,d.

\begin{acknowledgements}
We acknowledge support through the XMM/INTEGRAL PRODEX contract (Belspo), from the Fonds de Recherche Scientifique (FRS/FNRS), and CONACYT grant.
\end{acknowledgements}


\end{document}